# Monazite-type SrCrO$_4$ under compression

J. Gleissner[1,†], D. Errandonea[1,*], A. Segura[1], J. Pellicer-Porres[1], M. A. Hakeem[2], J. Proctor[2], S. V. Raju[3], R.S. Kumar[4], P. Rodríguez-Hernández[5], A. Muñoz[5], S. Lopez-Moreno[6], and M. Bettinelli[7]

[1]Departamento de Física Aplicada-ICMUV, MALTA Consolider Team, Universidad de Valencia, Edificio de Investigación, C/Dr. Moliner 50, Burjassot, 46100 Valencia, Spain

[2]School of Computing, Science, and Engineering, University of Salford, Manchester M5 4WT, UK

[3]CESMEC, Dept. Mechanical Engineering, Florida International University, Miami, FL, USA

[4]High Pressure Science and Engineering Center, Department of Physics, University of Nevada Las Vegas

[5]Departamento de Física, Instituto de Materiales y Nanotecnología, and MALTA Consolider Team, Universidad de La Laguna, La Laguna, E-38205 Tenerife, Spain

[6]CONACYT – Centre for Corrosion Research, Univeristy of Campeche, Av. Heroe de Nacozari 480, Campeche 24070, Mexico

[7]Luminescent Materials Laboratory, Department of Biotechnology, University of Verona and INSTM, UdR Verona, Strada Le Gracie 15, 37134 Verona, Italy

**Abstract:** We report a high-pressure study of monoclinic monazite-type SrCrO$_4$ up to 26 GPa. Therein we combined x-ray diffraction, Raman and optical-absorption measurements with *ab initio* calculations, to find a pressure-induced structural phase transition of SrCrO$_4$ near 8–9 GPa. Evidence of a second phase transition was observed at 10–13 GPa. The crystal structures of the high-pressure phases were assigned to the tetragonal scheelite-type and monoclinic AgMnO$_4$-type structures. Both transitions produce drastic changes in the electronic band gap and phonon spectrum of SrCrO$_4$. We determined the pressure evolution of the band gap for the low- and high-pressure phases as well as the frequencies and pressure dependences of the Raman-active modes. In all three phases most Raman modes harden under compression; however the presence of low-frequency modes which gradually soften is also detected. In monazite-type SrCrO$_4$, the band gap blue-shifts under compression, but the transition to the scheelite phase causes an abrupt decrease of the band gap in SrCrO$_4$. Calculations showed good agreement with experiments and were used to better understand the experimental results. From x-ray diffraction studies and calculations we determined the pressure dependence of the unit-cell parameters of the different phases and their ambient-temperature equations of state. The results are compared with the high-pressure behavior of other monazites, in particular PbCrO$_4$. A comparison of the high-pressure behavior of the electronic properties of SrCrO$_4$ (SrWO$_4$) and PbCrO$_4$ (PbWO$_4$) will also be made. Finally, the possible occurrence of a third structural phase transition is discussed.

† ERASMUS student from Imperial College London, London SW7 2AZ, United Kingdom

* Corresponding author, email: daniel.errandonea@uv.es

# I. Introduction

Photocatalytic materials which respond to ultra-violet (UV) and visible (VIS) light can be used in a wide variety of environmental applications [1]. As a consequence, they have received much attention in recent years. In particular, progress has been made thanks to the development of chromium-based compounds [1]. Among them, lead chromate ($PbCrO_4$) and strontium chromate ($SrCrO_4$) are the most studied materials due to their unique properties [2 - 5]. The crystal structures of these ternary oxides have been determined accurately [6], both being assigned to a monazite-type structure (space group $P2_1/n$, $Z = 4$). A schematic view of the monazite structure is given in Fig. 1. The structural arrangement is based on the nine-fold coordination of the Pb (Sr) cation and the fourfold coordination of the Cr cation. The ambient-pressure lattice vibrations and electronic band structures of $PbCrO_4$ and $SrCrO_4$ have already been studied too [7].

During the last decade, high pressure (HP) has been shown to be an efficient tool for improving the understanding of the physical properties of ternary oxides [8 – 15]. In particular, numerous monazite-type oxides have already been the subject of HP studies [16 - 19], which have concentrated mostly on phosphates and vanadates. Among the chromates, monazite-type $PbCrO_4$ is known to have quite an interesting high-pressure behavior [20, 21], undergoing several pressure-induced phase transitions. These transitions have important consequences in the electronic properties, modifying the electronic band gap from 2.3 eV at ambient pressure to 0.8 eV at 20 GPa [22]. To the best of our knowledge, and in contrast to $PbCrO_4$ and other monazite-type oxides, no HP studies of $SrCrO_4$ are available in the literature.

Here we will report a combined experimental and theoretical study of $SrCrO_4$ under compression. X-ray diffraction (XRD), Raman spectroscopy, and optical-absorption experiments have been carried out up to 26 GPa, which are complemented by a*b initio* calculations. We will report evidence of the existence of at least two phase transitions and propose crystal structures for the HP phases. The transitions have important consequences on the physical properties of $SrCrO_4$, which will be discussed in detail. The pressure-dependences of unit-cell parameters, Raman and infrared (IR) modes, and the electronic band gap will be also reported for the different phases. Moreover, a comparison of the high-pressure behavior of SrCrO$_4$ and related ternary oxides will be presented. The reported studies have enabled us to improve the understanding of the HP properties of SrCrO$_4$ and related compounds.

## II. Experimental details

SrCrO$_4$ in powder form was prepared by precipitation adding 50 ml of a 1 M Sr(NO$_3$)$_2$ solution to 50 ml of a 1 M K$_2$CrO$_4$ solution. Single crystals were grown using a ternary flux system composed of NaCl, KCl, and CsCl, as described in Ref. 23. The weight composition of the mixture was NaCl (24.8%), KCl (26.4%), CsCl (41.3%) and SrCrO$_4$ (7.5%). The starting reagents were mixed, placed in a platinum crucible with a tight-fitting lid, and kept for 12 h at 620 °C in a horizontal furnace under air atmosphere. The melt was slowly cooled in three steps: first to 530 °C with a temperature gradient of -1.5 °C/h, then to 450 °C at -2 °C/h, and finally to ambient temperature at -50 °C/h. The crystals were separated by careful dissolution of the flux in deionized water. Yellow single crystals of about 1 x 1 x 1 mm$^3$ were obtained. The purity of the synthesized material was confirmed by Energy-dispersive x-ray spectroscopy carried out in a transmission-electron microscope (TEM) operated at 200 KeV at the SC-SIE, Universitat de Valencia. By means of powder XRD measurements, using Cu K$_\alpha$ radiation, it was verified that the samples were single-phased and presented the monazite-type structure ($P2_1/n$). The unit-cell parameters were determined to be $a$ = 7.065(7) Å, $b$ = 7.376(7) Å, $c$ = 6.741(7) Å, and β = 103.1(1)°, in very good agreement with values reported in the literature [6, 24, 25, 26].

High-pressure powder XRD measurements were performed using a membrane diamond-anvil cell (DAC) and a 4:1 methanol-ethanol mixture as pressure-transmitting medium [27, 28]. The experiments were performed in the angle dispersive geometry with a symmetric-type DAC. The micron-sized powder, used throughout the experiments, was obtained by grinding single-crystals with pestle and mortar. The culet size of the diamond anvils was 400 μm and rhenium served as the gasket material. The gasket was pre-indented to a thickness of 50 μm and a hole with a diameter of 130 μm was drilled in its center to form a pressure chamber. Special caution was taken during the sample loading to avoid sample bridging between the diamond anvils [29]. Pressure was determined using the ruby scale [30]. Experiments were carried out at the beam line 12.2.2 of the Advanced Light Source, Lawrence Berkeley National Laboratory [31] with a MAR345 detector. Monochromatic x-rays with a wavelength of 0.4949 Å were used for the experiments and the FIT2D software [32] was employed to calibrate sample to detector distance and detector tilt, as well as to integrate the two-dimensional diffraction images to standard one dimensional intensity versus $2\theta$ plots. The structural analysis was performed with the GSAS and Powdercell software packages [33, 34].

Four independent Raman experiment were performed. Two runs were carried out using one set-up and the other two with a different set-up. In the first two runs, HP Raman spectra were collected in the backscattering geometry using a 632.8 nm He-Ne laser, a single spectrometer (Jobin−Yvon TRH1000), an edge filter and a thermoelectric-cooled multichannel CCD detector (Horiba Synapse). The set-up was calibrated using plasma lines of the He-Ne laser. The other two runs were carried out in the backscattering geometry using a 532 nm diode laser. In these experiments, the scattered light was analyzed with a Jobin-Yvon Raman system equipped with a single spectrograph, an edge filter, and an air-cooled multichannel CCD detector (iDus 420). This set-up was calibrated using the Raman lines of Si and diamond. In all the experiments, a laser power of less than 20 mW before the DAC was used to avoid sample heating and the spectral resolution of the system was below 2 $cm^{-1}$. The experiments were carried out using 10-μm-thick single crystals of $SrCrO_4$ which were loaded either in a symmetric DAC or in a membrane DAC. In both cases we used ultralow fluorescence diamond anvils (300 – 500 μm size) and either inconel or stainless steel gaskets. The gaskets were pre-indented to a thickness of 40-50 μm and a hole of 100-200 μm was used as the pressure chamber. As pressure medium we used either a 16:3:1 methanol-ethanol-water mixture (MEW) or nitrogen [27]. The four experiments gave similar results. Pressure was determined using the ruby scale [30].

For optical absorption studies, we used 10-µm-thick parallel face crystals, which were cleaved from the larger single crystals. Measurements in the UV-VIS-near-infrared (NIR) range were made with an optical setup that consisted of deuterium and halogen lamps integrated in the DH-2000 light-source from Ocean Optics, fused silica lenses, reflecting optics objectives, and an Ocean Optics USB2000 UV−VIS−NIR spectrometer [35, 36]. The absorption spectra were obtained from the transmittance spectra of the sample, which were recorded using the sample-in, sample-out method [37, 38]. For these experiments we used a membrane DAC equipped with 500 μm culet type IIA diamonds. The pressure chamber consisted in a 200 μm diameter hole drilled in a 45 μm thick inconel gasket. Ruby fluorescence was used as pressure standard [30] and a mixture of methanol-ethanol-water (16:3:1) was employed as the pressure-transmitting medium [27]. Three independent experiments, which involve three different samples, were carried out. At ambient pressure, we found that variation in the optical band gap with crystal orientation is minimum and comparable to the accuracy of the measurements. Based upon this fact, our HP results neglect effects of crystal orientation on the band-gap values.

We would like to add here that in all the experiments described above the ruby lines showed a reasonable full-width at half maximum indicating than even at pressure where the pressure media were not quasi-hydrostatic deviatoric stresses were small. In addition, we also took care that the sample occupied only a small fraction of the pressure chamber and the gasket never distorted during experiments. We think these facts that the phase transition we will reported here are intrinsic to the application of pressure.

### III. Computational details

*Ab initio* simulations of SrCrO$_4$ under pressure were performed within the framework of Density-Functional Theory (DFT) [39], as implemented in the Vienna *ab initio* simulation package (VASP) [40]. The pseudopotential with the projector augmented wave scheme (PAW) [41] was employed to describe the atomic species. Due to the presence of oxygen atoms, the set of plane waves was developed up to a kinetic energy cut off of 520 eV, in order to obtain accurate results. The exchange-correlation energy was described in the generalized-gradient approximation (GGA) with the Perdew-Burke-Ernzenhof prescription for solids (PBEsol) [42]. To carry out integrations over the Brillouin zone (BZ), dense meshes of Monkhorst-Pack special k-points [43] appropriate to each structure were used. The convergence achieved in energy was better than 1 meV per formula unit. At selected volumes, and for each structure considered, the lattice parameters and atomic positions were fully optimized trough the calculation of forces on atoms and the stress tensor. In the optimized structures, the forces on atoms were less than 0.004 eV/ Å and the deviations of stress tensor components from the diagonal hydrostatic form were lower than 0.1 GPa. From the set of energy (E), volume (V), and pressure (P) data, the enthalpy (H) as a function of P was obtained and the relative stability between the different phases was analyzed. DFT is a well-tested method, which accurately describes the relative phase stability and the properties of semiconductors under high pressure [44]. The electronic band structure along high symmetry directions in the Brillouin zone and the density of states were also calculated.

The direct force-constant method [45] was employed to study the lattice vibrations. High-pressure lattice dynamic calculations were carried out at the zone center ($\Gamma$ point) of the BZ. The diagonalization of the dynamical matrix provided the frequency of the Raman and infrared modes. The construction of the dynamical matrix at the $\Gamma$ point required highly accurate calculations of the forces which appear on the atoms when small displacements from their equilibrium configuration are considered. From the calculations, symmetry and eigenvectors of the vibration modes of the considered structures at the $\Gamma$ point are also identified. The mechanical and phonon stability of the different phases was also evaluated.

## IV. Results and discussion

### A. Effects of pressure on the crystal structure

From our experiments we found evidence for at least two pressure-induced phase transitions in SrCrO$_4$. Since the interpretation of the experiments will be based upon *ab initio* calculations, we will first present the results of our theoretical study on the structural stability of SrCrO$_4$ at high pressures. In the calculations we have taken into consideration previous results obtained for monazite-type oxides [16, 17, 21, 46] and also candidate HP structures predicted by the packing-efficiency criterion proposed by Bastide [15]. We have studied the relative stability of several candidate HP structures using the calculation method outlined in the previous section. In Fig. 2 we report the difference of enthalpy (taking monazite as reference) of the structures that we found to be thermodynamically competitive with monazite. At ambient pressure, monazite is the most stable structure of SrCrO$_4$. The calculated structure is reported in Table I where it is compared with experiments previously reported by us [7]. The agreement is excellent. As can be seen in Fig. 2, at 7 GPa a transition from the monoclinic monazite-type structure to a tetragonal scheelite-type structure (space group *I*4$_1$/*a*, Z = 4) is suggested by the calculations. At 14 GPa a subsequent transition to a monoclinic AgMnO$_4$-type structure (space group *P*2$_1$/*n*, Z = 4) is predicted by the calculations. Details of the calculated HP crystal structures are given in Tables II and III. A schematic view of both HP structures can be seen in Fig. 1. The two phase transitions proposed by calculations are in agreement with the experiments reported below. According with calculations the predicted phase transitions are not caused by mechanical or phonon instabilities being probably ions interactions the trigger factor of the detected phase transitions.

In order to confirm the existence of pressure-induced phase transitions in SrCrO$_4$ we performed ambient-temperature XRD measurements. A selection of diffraction patterns at different pressures is given in Fig. 3. We found that up to 6.8 GPa the XRD patterns can be Rietveld refined assuming the monazite structure. In the figure we show the results of the experiments carried out at 1.2 GPa and 6.8 GPa together with Rietveld refined profiles and the residuals, which support the identification of the monazite structure. At 1.2 GPa, the goodness-of-fit parameters are: $R_P$ = 4.47%, $R_{WP}$ = 6.76%, and $\chi^2$ = 1.64. Similar figures of merit were obtained at all pressures for the monazite structure; e.g. at 6.8 GPa $R_P$ = 4.94%, $R_{WP}$ = 7.22%, and $\chi^2$ = 1.84. Before discussing the phase transitions induced by pressure in $SrCrO_4$, we would like to mention that when comparing the XRD pattern measured at 6.8 GPa with the one measured at 1.2 GPa, it can be seen that several Bragg peaks split as pressure increases. This is clearly seen in the figure for the (012) and (-112) peaks which are labelled accordingly. This fact suggests a non-isotropic compression of monazite $SrCrO_4$, which will be discussed after presenting the evidence of the observed phase transitions.

When increasing the pressure from 6.8 GPa to 9.4 GPa very noticeable changes take place in the XRD pattern. These changes are consistent with the occurrence of a phase transition at 7.5 GPa as predicted by our calculations. The reduction in the number on Bragg reflections presents a strong indication for a symmetry increase in the crystal structure. In particular, the XRD patterns we measured at 9.4 and 11.2 GPa can be indexed assuming the tetragonal scheelite-type structure. In Fig. 3 we show the results of a Rietveld refinement carried out for the XRD pattern measured at 9.4 GPa. The residuals are small, which indicates that scheelite is a suitable structural model for the crystal structure of the HP phase. The structural information of the scheelite-type phase is given in Table II. The agreement between calculations is good, not only for the unit-cell parameters, but also for the atomic positions of the oxygen atoms (The positions of Sr and Cr are fixed by the symmetry of the structure). The goodness-of-fit parameters of the refinement shown in Fig. 3 for the scheelite structure are: $R_P$ = 5.74%, $R_{WP}$ = 7.94%, and $\chi^2$ = 2.12.

When increasing the pressure from 11.2 GPa to 13.3 GPa, we found evidence for a second phase transition, which agrees with the 12 GPa transition pressure found by calculations for the scheelite-$AgMnO_4$-type transition. In particular, the XRD patterns measured from 13.3 GPa up to 18.9 GPa can be properly refined assuming the $AgMnO_4$-type structure. The results of the refinement performed at 13.3 GPa and 16.7 GPa are shown in Fig. 3. The refinements indicate that the AgMnO$_4$-type structure can be assigned to the second HP phase of SrCrO$_4$. The goodness-of-fit parameters of the refinement for the XRD pattern measured at 13.3 GPa are: $R_P$ = 5.97%, $R_{WP}$ = 8.02%, and $\chi^2$ = 2.19. The obtained unit-cell parameters and atomic positions are given in Table III. The agreement between experiments and calculations is quite good, which makes us confident about the structural assignment made for the second HP phase of SrCrO$_4$. We would like to highlight the fact that the coordination of Cr is not affected during the monazite-scheelite-AgMnO$_4$-type structural sequence, Cr being coordinated by four oxygens forming a regular (or nearly regular) CrO$_4$ tetrahedron in all three structures. However the Sr coordination is modified, becoming the coordination number of Sr ten the AgMnO$_4$-type structure.

After a subsequent compression step from 18.9 GPa to 20.4 GPa we observed important changes in the XRD pattern (see Fig. 3). These changes indicate that possibly a third phase transition is taking place. Unfortunately, the low quality of the XRD pattern measured at 20.4 GPa does not allow the identification of the crystal structure of the third HP phase, which we will name phase IV. In fact we cannot exclude the phase coexistence of phase IV and the AgMnO$_4$ phase at 20.4 GPa. With the aim of trying to clarify this last hypothesis we increase the pressure in two steps up to 24 GPa. However, the diffraction peaks broaden, which precludes any sound structural identification. Thus, from our XRD experiments we can only state that the onset of a third phase transition takes place between 18.9 and 20.4 GPa. This conclusion is supported by our Raman experiments, as we will comment below. The identification of the crystal structure of phase IV remains an open issue for future studies. Before concluding this part of the discussion we would like to mention that upon a rapid decompression from 24 GPa to 0.1 GPa the crystal structure of the low-pressure monazite phase was recovered. This is shown in Fig. 3. There it can be seen that the XRD pattern measured after decompression at 0.1 GPa is quite similar to the one measured at 1.2 GPa during compression.

From the XRD experiments and the calculations we have determined the pressure dependence of the unit-cell parameters. The results are shown in Fig. 4 where the symbols and lines represent the experimental results and calculations, respectively. We found that the low-pressure phase is slightly more compressible than the two HP phases. This is consistent with the volume reduction associated with each phase transition. In Fig. 4, it can be seen that in both phase transitions there is a discontinuity in the volume, which is larger than the uncertainty of the volume determination. While *ab initio* calculations correctly predict the overall volume change between the monazite and AgMnO$_4$ phases, there is a discrepancy between the predicted and measured relative volume changes at the monazite-scheelite (-2% theory versus -4% experiment) and scheelite-AgMnO$_4$ (-4% theory versus -2% experiment). This small discrepancy is most probably due to the narrow pressure range through which the scheelite phase is observed, which certainly limits the accuracy in the determination of the scheelite phase equation of state (EOS). From the volume discontinuities observed in the transitions, it can be stated that both structural changes are first-order transitions. Regarding the reduction of the lattice parameters, we can conclude that the compression in the low-pressure monazite phase is anisotropic, with the *a*-axis being the most compressible axis. In addition, it can be seen that the β angle is reduced by compression; approximately 0.2º per GPa. The observed behavior of monazite SrCrO$_4$ is qualitatively similar to that of other monazites [16 – 20, 47]. In contrast to the low-pressure phase, the compression in the scheelite and AgMnO$_4$ structures is nearly isotropic, with the β angle of the last phase being only slightly reduced by compression.

From the pressure dependence of the unit-cell parameters we determined the pressure-volume equation of state (EOS) for the three phases of SrCrO$_4$ and their compressibility tensor. Since we have a few experimental data points for each phase and calculations and experiments qualitatively give similar pressure dependence for the unit-cell volume, we have used the calculations to quantitatively describe the compression of the different phases. We found that for the three phases, the pressure dependence of the volume can be well described by a third-order Birch–Murnaghan

EOS [48]. The obtained EOS parameters are summarized in Table IV. In the table $V_0$ is the unit-cell volume at ambient pressure, $B_0$ the bulk modulus, and $B_0$' its pressure derivative. The use of a third-order EOS was based upon an analysis of the dependence of the normalized pressure on the Eulerian strain [49]. Among the different polymorphs of $SrCrO_4$, monazite is the one with the smallest bulk modulus. Regarding the compressibility tensor, in a monoclinic structure this tensor has four independent components $\beta_{11}$, $\beta_{22}$, $\beta_{33}$, and $\beta_{13}$. The analytical expressions of them can be found in Ref. 50. In our monoclinic structures (where $b$ is the unique crystallographic axis) $\beta_{22}$ and $\beta_{33}$ are the compressibilities of the $b$ and $c$ axes, respectively. On the other hand $\beta_{11}$ corresponds to the compressibility in the direction perpendicular to the $b$-$c$ plane and $\beta_{13}$ describes the change of the shape of the plane perpendicular to the unique crystallographic axis. In the case of the tetragonal scheelite structure, given the symmetry of the crystal, $\beta_{11} = \beta_{22}$ and $\beta_{13} = 0$. The values obtained for $\beta_{11}$, $\beta_{22}$, $\beta_{33}$, and $\beta_{13}$ for the three phases are given in Table IV. In the table it can be confirmed that the compression of the low-pressure phase is non-isotropic. This is indicated by the fact that $\beta_{33}$ is more than 20% larger than $\beta_{11}$ and $\beta_{22}$. In contrast in the other two phases the diagonal components of the tensor have values that differ by less than 10%. In the case of the $AgMnO_4$ structure, $\beta_{13}$ is quite small compared to the same parameter in the monazite structure. This fact indicates that the shape of the unit-cell of $AgMnO_4$ is basically not modified by compression. This and the value of the β angle, which is close to 90°, suggest that as a first approximation this HP phase behaves as a quasi-orthorhombic structure, which resembles a distorted-barite structure.

### B. Raman spectroscopy

We will now present Raman-spectroscopic evidence on the pressure-driven transitions in $SrCrO_4$. We have previously reported the ambient pressure Raman spectrum of monazite $SrCrO_4$ as well as the mode frequency and assignment of the Raman-active modes [7]. The modes have been identified as internal stretching (high frequency) and bending (intermediate frequency) modes of the $CrO_4$ tetrahedron and external modes (low frequency), which involve movements of both the $Sr^{2+}$ and $CrO_4^{2-}$ ions [7]. Previously, thirty of the thirty-six expected modes ($\Gamma = 18A_g + 18B_g$) were measured at ambient pressure (outside the DAC) [7]. In this paper, thirty-three modes have been detected. The wavenumbers of these modes are given in Table V. In Figures 5 and 6 we show Raman spectra measured at HP using MEW as the pressure medium and a 632.8 nm He-Ne laser excitation. To facilitate the mode identification we have divided the Raman spectra into three regions. They correspond to the external and internal (bending and stretching) modes, which have very different intensities. The spectra shown in Figure 5 correspond to measurements carried out up to 8.2 GPa. All the Raman spectra shown in this figure resemble the ambient pressure Raman spectrum of monazite $SrCrO_4$. In the HP experiments, we have been able to identify a maximum of twenty-six modes of the monazite phase. They are identified by ticks in the figure for the spectrum measured at 1.1 GPa. The weakest modes observed at ambient pressure (outside the DAC) were not observed at HP because the presence of the diamond anvils increases the background level thereby decreasing the signal-to-noise ratio [51]. Most of the modes could be followed up to 8.2 GPa. Up to this pressure the Raman spectra can be unambiguously assigned to the monazite phase. Since not all modes are equally affected by pressure, a tendency of some of them to merge under compression was observed. Clear evidence of three phonons crossing over other Raman modes was also observed. Two of the phonon crossovers occur for external modes, and one for internal stretching modes. Most of the modes harden under compression. Only two modes have negative pressure coefficients according to experiments. The frequencies ($\omega$) and pressure coefficients ($d\omega/dP$) of the different modes are summarized in Table V where they are compared with our theoretical calculations. The Grüneisen parameters ($\gamma$), which provide a dimensionless representation of the response to compression, are also included. The pressure dependence of the Raman frequencies is presented in Figure 7. In the table we have included a column to show the relative difference between experimental and theoretical frequencies, $R_\omega$, as defined in Ref. 52. For most modes $R_\omega$ is smaller than 5% and in many modes even smaller than 1%, which illustrates the excellent agreement between calculations and experiments. Regarding the pressure dependence of the modes there is a qualitative agreement between calculations and experiments. We observed that the behavior observed in monazite-type $SrCrO_4$ is qualitatively similar to that of isomorphic $PbCrO_4$ [20]. A remarkable feature is the presence of two external modes below 100 cm$^{-1}$ which have negative pressure coefficients. Theory predicts the existence of a third mode whose frequency decreases under compression. However, this mode is not detected in our experiments because it is expected to be below the low-frequency limit of the Raman set-up. The presence of these modes is apparently a typical feature of monazites since it has been also detected in $PbCrO_4$ [20] and $LaVO_4$ [46]. The presence of such modes might be correlated with a weakening of the restoring force against the corresponding deformation associated to the phonon mode, probably marking the existence of a collective instability that tends to make the crystal structure unstable. This fact is consistent with the finding of a phase transition at relatively low pressures as it was found in $SrCrO_4$. However, since the wavenumber of the modes with negative pressure coefficients never reaches zero, they are not classical soft-modes, as those observed in a second-order displacive transition [53, 54]. Another feature to be remarked upon is the tendency of the external modes of monazite $SrCrO_4$ to have larger Grüneisen parameters than the internal modes. The same trend has been observed before, not only in monazite $PbCrO_4$ [20], but also in barite-type $BaCrO_4$ [55].

When increasing the pressure from 8.2 GPa to 8.9 GPa very important changes take place in the Raman spectrum. In the spectrum measured at 8.9 GPa only eleven modes can be identified. The observed changes indicate the occurrence of a phase transition. The transition pressure is consistent with the monazite-scheelite transition pressure obtained from XRD and calculations. The mode distribution of phonons in the Raman spectrum is very similar to that of most scheelite oxides [56]. The Raman spectrum of the scheelite structure has thirteen Raman-active modes ($\Gamma = 3A_g + 5B_g + 5E_g$) [56], but in our case we detected only eleven. They are identified by ticks in Figure 5 for the spectrum measured at 8.9 GPa. Regarding the undetected modes, one of the modes not detected (with a wavenumber of 175 cm$^{-1}$, according to calculations) is usually very weak [56]. The other mode is likely not detected due to the overlap of two modes at 375 cm$^{-1}$. Confirmation of the assignment of the Raman spectrum measured at 8.9 GPa to the scheelite phase comes from *ab initio* calculations. In Table VI we report the Raman frequencies determined from experiments and calculations. All frequencies agree within 5%, supporting the finding that the measured Raman spectrum can be assigned to the HP scheelite structure. Calculations also provide the mode assignment which is given in the table. A typical feature of the scheelite Raman spectrum is the presence of three strong modes in the high-frequency region, which are indeed present in the spectra we assigned to the scheelite structure. The modes are internal stretching modes of the $CrO_4$ tetrahedron and are separated by a large phonon gap from the rest of the modes (see Table VI). When increasing the pressure we observed the scheelite phase, as a single phase, over a reduced pressure range because of the onset of a second phase transition at 9.7 GPa (see discussion below). However, the most intense peaks of the scheelite phase can be detected up to 11.7 GPa. The phonon frequencies as a function of pressure are shown in Figure 7. From these results we estimate the pressure coefficient of each phonon. The coefficients are shown in Table VI where they are compared to calculations. The agreement for pressure coefficients is not as good as for the frequencies. However, differences are comparable with the discrepancy observed between theory and calculations for the HP phases of related oxides [57]. The mode with the largest discrepancy in the pressure dependence of the frequency is the low frequency $B_g$ mode at 127 cm$^{-1}$. In spite of these facts, both methods gave a qualitatively similar picture, suggesting that in scheelite $SrCrO_4$, as is also the case in the low-pressure monazite phase, the external modes ($\omega < 375$ cm$^{-1}$) are the modes with the largest Grüneisen parameters. In addition, as in the monazite phase, in scheelite $SrCrO_4$ there is also a phonon with a negative pressure coefficient. This is the phonon with the lowest frequency (see Table VI). The presence of a mode with such a behavior is a distinctive feature of scheelite-structured oxides [58].

As we commented above, at 9.7 GPa, additional Raman modes appear in the spectra suggesting the onset of a second phase transition. We found evidence for the coexistence of the scheelite phase with the new HP phase up to 11.7 GPa. The modes of the new phase gradually become stronger while the scheelite modes lose intensity. The coexistence of the two phases can be a consequence of the fact that the proposed structural phase transition is an order-disorder transition. At 12.2 GPa the scheelite modes have completely vanished. The existence of this second transition is in agreement with the conclusions drawn from our XRD experiments and calculations. It is noticeable that the new HP phase has many more Raman modes than scheelite, which is consistent with the scheelite-$AgMnO_4$ transition. From 12.2 GPa to 14.7 GPa we did not observe any qualitative change in the Raman spectrum. We will show below that the calculated Raman frequencies for the $AgMnO_4$ phase agree reasonably well with the modes we identified in the experiments. At 15.7 GPa we observe the appearance of several additional Raman modes and the disappearance of some of the modes of the $AgMnO_4$ phase. These changes suggest another transition to a phase which we will refer to as phase IV. As we mentioned above this transition was detected by XRD at 20.4 GPa and the identification of the crystal structure of phase IV is beyond the scope of this work. We would like to mention here that the same transition was detected when using nitrogen as the pressure medium at 19.5 GPa. The difference in the transition pressures for the $AgMnO_4$-phase IV transition can be caused by the use of different pressure media, which have a different hydrostatic pressure limit [27], therefore influencing the transition pressures of compounds like $SrCrO_4$ [59, 60]. Before discussing the Raman modes of $AgMnO_4$-type $SrCrO_4$ in more detail, we would like to add that we observed that phase IV remains stable up to 26 GPa.

From factor group analysis, it can be established that the $AgMnO_4$ structure presents 36 Raman-active phonons ($\Gamma$ = 18Ag + 18Bg), exactly as the monazite phase. The expected number of Raman modes is consistent with the changes we observed in the Raman spectra near the scheelite-$AgMnO_4$ transition pressure. The calculated wavenumbers and mode assignment of all Raman-active modes for the $AgMnO_4$-type structure are given in Table VII. We have eighteen low-frequency lattice modes plus ten internal bending modes and eight internal stretching modes of the $CrO_4$ tetrahedron. In the experiments we have also detected thirty six modes. A correlation can be established between calculated and measured frequencies for the eighteen low-frequency lattice modes. However, the internal modes cannot be fully correlated. In fact, for the intermediate frequency range (340 < ω < 450 cm$^{-1}$) we have measured only eight modes, whereas calculations predict ten modes. The discrepancy could be caused by the fact that calculations predict that two couples of $A_g/B_g$ are very close in frequency. This added to the fact that modes broaden and lost intensity as pressure increases, could justify the detection of only eight Raman modes instead of the expected ten modes. In the high-frequency region the opposite behavior is found. Eight modes are predicted by theory while we observed ten modes in the experiments. The two extra modes could be overtones of the low-frequency modes or be induced by a disorder in the crystal structure as previously observed when disorder is induced in related oxides [60, 61]. In summary, however, we can state that calculations and experiments show a qualitative overall agreement on the Raman spectrum of the second HP phase of $SrCrO_4$, indicating that the $AgMnO_4$-type structure we determined from XRD and calculations gives a good model to explain the Raman spectrum of the second HP phase. Regarding the pressure dependence of the Raman modes, the agreement between experiments and calculations is good, even better than for the scheelite phase. The main difference between the $AgMnO_4$ and the other two phases is that in the $AgMnO_4$ phase the external and internal modes have similar Grüneisen parameters. This feature could be accounted for by the fact that, in the high pressure phase, external bonds become stronger and less compressible than in the ambient pressure phase. Another point to note is that the $AgMnO_4$ phase has two phonons with negative pressure coefficients, which are the two lowest frequency modes.

### C. Optical absorption and band structure

Figure 8 shows a selection of optical-absorption spectra measured at different pressures. From the parabolic dependence of the absorption coefficient on the photon energy, it can be concluded that $SrCrO_4$ is an indirect band-gap material with band-gap energy ($E_g$) of 2.45(5) eV. Band-structure calculations confirm that $SrCrO_4$ has an indirect band gap $E_g$ = 2.67 eV (from Γ to Γ-Z). The agreement between calculations and experiments is good with theory overestimating $E_g$ by just 9%. As pressure increases, the absorption edge shifts towards higher energy, resulting in a change of SrCrO$_4$ crystal color from orange to orange-yellow. At 8.3 GPa we observed an abrupt shift of the absorption edge towards low energies. At this pressure the crystal of SrCrO$_4$ changes it color becoming orange-red. This abrupt change in the optical properties correlates well with the occurrence of the monazite-scheelite transition. Upon further compression, there is a blue-shift of the absorption spectrum of SrCrO$_4$. From the optical-absorption measurements the pressure dependence of E$_g$ up to 15 GPa was determined, as shown in Figure 9. For the low-pressure phase, we found a gradual increase of Eg under compression. Assuming there is a linear relation between E$_g$ and pressure, we determined dE$_g$/dP = 17(5) meV/GPa. This pressure coefficient contrasts with the pressure coefficient determined for the band gap of monazite PbCrO$_4$ (-46 meV/GPa) [22]. Note that the same differences are found when comparing the pressure effect on the band gap of SrWO$_4$ (dE$_g$/dP = 3.7 meV/GPa) and PbWO$_4$ (dE$_g$/dP = -61 meV/GPa) [35]. An explanation to the observed difference comes from band structure calculations. Our calculations indicate that in both compounds, the upper part of the valence band is dominated by O 2p states. In contrast, the lower part of the conduction band is composed primarily of electronic states associated with the Cr 3d and O 2p states. On the other hand, in SrCrO$_4$ the Sr states are completely empty near the Fermi level. Thus, they do not have any influence on the bandgap energy. However, in PbCrO$_4$, there is a contribution of Pb 6s electrons to the top of the valence band and of Pb 6p states to the bottom of the conduction band. As a consequence, E$_g$ is smaller in PbCrO$_4$ (2.25 eV) than in SrCrO$_4$ (2.45 eV). Another consequence is the different behavior of E$_g$ with pressure in both compounds. From our calculations, we found that under compression in SrCrO$_4$ Cr 3d states move faster towards higher energies than O 2p states, leading to the small opening of the band gap we observed in the experiments. In contrast, in PbCrO$_4$, the top of the valence band shifts toward high energies faster than the bottom of the conduction band. This is a consequence of the separation between Pb bonding and anti-bonding states becoming enlarged under compression. This fact enhances the displacement towards higher energies of the top of the valence band, but reduces the displacement of the bottom of the conduction band.

At 8.3 GPa an abrupt decrease of 0.2 eV in $E_g$ was observed (see Figure 9). This pressure corresponds to the monazite-scheelite transition we described above. We carried out band-structure calculations for scheelite-type $SrCrO_4$. Calculations indicate that in the scheelite structure $SrCrO_4$ is a direct band-gap material with $E_g = 2.25$ eV (at 8.3 GPa) with the band gap located at the $\Gamma$ point of the Brilloun zone. This value of Eg is 6% smaller than the experimental value determined for the scheelite phase; $E_g = 2.40(5)$ eV. The band-gap collapse determined from calculations is 0.4 eV. In summary, the changes observed in the optical properties of $SrCrO_4$ at 8.3 GPa are consistent with the structural sequence found in our structural and vibrational studies. Upon further compression in the scheelite phase we observed that $E_g$ linearly increases with pressure with $dE_g/dP = 16(5)$ meV/GPa, which is nearly identical to the behavior of the low-pressure phase. This result is consistent with the fact that in both structures the valence and conduction bands near the band gap are dominated by molecular orbitals associated with the $CrO_4^{-2}$ ions, and the fact that the $CrO_4$ tetrahedron undergoes a similar compression in both structures.

At 10.2 GPa a change in the pressure dependence of $E_g$ was found. This change is consistent with the occurrence of a second phase transition, as discussed in the previous sections based upon our structural and vibrational studies. For the second HP phase we determined $dE_g/dP = 4(2)$ meV/GPa; which indicates that the band gap is less sensitive to pressure in the $AgMnO_4$-type phase. The decrease of $dE_g/dP$ in this phase is consistent with the fact that this is the least compressible structure among the three structures reported for $SrCrO_4$. For the second HP phase the experimentally determined bandgap is $E_g = 2.46(5)$ eV at 14.5 GPa, in excellent agreement with *ab initio* band structure calculations yielding $E_g = 2.45$ eV for the $AgMnO_4$-type phase. According to our calculation in this phase $SrCrO_4$ is an indirect band gap material, with the top of the valence band at the Y point of the Brillouin zone and the bottom of the conduction band at the $\Gamma$ point.

## V. Summary

We have performed HP XRD, Raman, and optical-absorption measurements as well as *ab initio* calculations on $SrCrO_4$. Changes in the crystal structure, lattice dynamics, and optical properties indicate the occurrence of at least two phase transitions. *Ab initio* calculations confirm the experimental findings and help to understand them. We have assigned a scheelite-type and a $AgMnO_4$-type structure to the two HP polymorphs found in $SrCrO_4$. The pressure dependence of unit-cell parameters, Raman modes, and band-gap energy is reported for the low-pressure monazite phase and the two HP phases of $SrCrO_4$. An assignment for the Raman modes is proposed based upon calculations. The reported results augment the understanding of the effects of pressure on the physical properties of ternary oxides. A comparison with the behavior of the band gap of $SrCrO_4$ and $PbCrO_4$ is presented and an explanation to their different HP behaviors is proposed. The fact that the electronic states near the Fermi level are mainly Cr 3d and O 2p states makes the band gap of monazite $SrCrO_4$ less sensitive to pressure than in $PbCrO_4$. This conclusion also explains the distinctive pressure behavior of the band gap of $SrWO_4$ and $PbWO_4$.

**Acknowledgments**

This work has been done under financial support from Spanish MINECO under projects MAT2013-46649-C4-1/3-P and MAT2015-71070-REDC. Supercomputer time has been provided by the Red Española de Supercomputación (RES) and the MALTA cluster. The authors thank the SCSIE from Universitat de Valencia for the technical support. S.M.L. thanks CONACYT from Mexico for financial support through the program Catedras for young researchers.

**Table I:** Structural parameters of the monazite structure ($P2_1/n$) at ambient pressure.

| | | | | | | | |
|---|---|---|---|---|---|---|---|
| Experiment | $a = 7.065(7)$ Å, $b = 7.376(7)$ Å, $c = 6.741(7)$ Å, $\beta = 103.1(1)°$ | | | | | | |
| Theory | $a = 7.0846$ Å, $b = 7.3843$ Å, $c = 6.6972$ Å, $\beta = 103.27°$ | | | | | | |

| Atom | site | Theory | | | Experiment | | |
|---|---|---|---|---|---|---|---|
| | | x | y | z | x | y | z |
| Sr | 4e | 0.2269 | 0.1574 | 0.3980 | 0.22813 | 0.15869 | 0.39806 |
| Cr | 4e | 0.1974 | 0.1651 | 0.8860 | 0.19769 | 0.16487 | 0.88691 |
| O$_1$ | 4e | 0.2583 | 0.0033 | 0.0597 | 0.2584 | 0.0055 | 0.0562 |
| O$_2$ | 4e | 0.1188 | 0.3393 | 0.0018 | 0.1201 | 0.3373 | 0.0024 |
| O3 | 4e | 0.0231 | 0.1002 | 0.6917 | 0.0256 | 0.1012 | 0.6981 |
| O$_4$ | 4e | 0.3798 | 0.2214 | 0.7848 | 0.3776 | 0.2179 | 0.7881 |

**Table II:** Structural parameters of the scheelite structure ($I4_1/a$) at 9.8 GPa (theory) and 9.4 GPa (experiments).

| Experiment | $a = 4.970(8)$ Å, $c = 11.844(7)$ Å |
|---|---|
| Theory | $a = 5.0105$ Å, $c = 11.8703$ Å |

| Atom | site | Theory | | | Experiment | | |
|---|---|---|---|---|---|---|---|
| | | x | y | z | x | y | z |
| Sr | 4b | 0 | 0.25 | 0.625 | 0 | 0.25 | 0.625 |
| Cr | 4a | 0 | 0.25 | 0.125 | 0 | 0.25 | 0.125 |
| O | 16f | 0.2454 | 0.1345 | 0.0471 | 0.2373(9) | 0.1126(9) | 0.0451(9) |

**Table III:** Structural parameters of the AgMnO$_4$ structure (*P2$_1$/n*) at 13.2 GPa (theory) and 13.3 GPa (experiments).

Experiment   $a = 6.680(8)$ Å, $b = 6.881(8)$ Å, $c = 6.118(8)$ Å, $\beta = 92.33(9)°$

Theory   $a = 6.6776$ Å, $b = 6.8970$ Å, $c = 6.1296$ Å, $\beta = 92.63°$

| Atom | site | Theory | | | Experiment | | |
|---|---|---|---|---|---|---|---|
| | | x | y | z | x | y | z |
| Sr | 4e | 0.3471 | 0.8639 | 0.2292 | 0.3563(1) | 0.8743(1) | 0.2265(1) |
| Cr | 4e | 0.3224 | 0.6335 | 0.7727 | 0.3227(1) | 0.6307(1) | 0.7620(1) |
| O$_1$ | 4e | 0.0052 | 0.0406 | 0.2683 | 0.0200(5) | 0.0217(5) | 0.2653(5) |
| O$_2$ | 4e | 0.1631 | 0.6127 | 0.9708 | 0.1606(5) | 0.6140(5) | 0.9742(5) |
| O3 | 4e | 0.2652 | 0.1541 | 0.9631 | 0.2777(5) | 0.1591(5) | 0.9600(5) |
| O$_4$ | 4e | 0.4492 | 0.8382 | 0.8232 | 0.4205(5) | 0.8460(5) | 0.8212(5) |

**Table IV:** EOS parameters and components of the compressibility tensor of the three different phases of SrCrO$_4$. These components have been calculated at 0 GPa for the monazite structure, at 8.2 GPa for the scheelite structure, and at 12.4 GPa for the AgMnO$_4$ structure.

|  | Monazite | Scheelite | AgMnO$_4$ |
|---|---|---|---|
| V$_0$ (Å$^3$) | 341(1) | 329(1) | 327(1) |
| B$_0$ (GPa) | 59(1) | 66(2) | 69(3) |
| B$_0$' | 4.9(5) | 4.8(5) | 4.5(5) |
| $\beta_{11}$ (10$^{-3}$ GPa$^{-1}$) | 6.77 | 3.06 | 2.63 |
| $\beta_{22}$ (10$^{-3}$ GPa$^{-1}$) | 5.53 | 3.06 | 2.56 |
| $\beta_{33}$ (10$^{-3}$ GPa$^{-1}$) | 5.12 | 2.72 | 2.79 |
| $\beta_{13}$ (10$^{-3}$ GPa$^{-1}$) | -2.25 | 0 | -0.11 |

**Table V:** Ambient pressure experimental and calculated wavenumbers (ω) for Raman modes of monazite-type SrCrO$_4$ (in cm$^{-1}$) including mode assignment. The pressure coefficients (dω/dP) are also reported (in cm$^{-1}$/GPa) as well as the Grüneisen parameters (γ). The relative difference between measured and calculated frequencies (R$_ω$) is also given (in %).

| Mode | Theory | | | Experiments | | | R$_ω$ |
|---|---|---|---|---|---|---|---|
| | ω | dω/dP | γ | ω | dω/dP | γ | |
| A$_g$ | 58.4  | -0.7 | -0.71 | ---- | ---- | ---- | ---- |
| B$_g$ | 64.8  | 2.2  | 2.00  | 67   | 1.7  | 1.50  | -3.3 |
| A$_g$ | 76.6  | -2.0 | -1.54 | 78   | -0.6 | -0.45 | -1.8 |
| A$_g$ | 91.7  | -1.0 | -0.64 | 89   | -0.1 | -0.07 | 3.0  |
| B$_g$ | 92.2  | 0.2  | 0.13  | 94   | 0.8  | 0.50  | -1.9 |
| A$_g$ | 106.6 | 1.9  | 1.05  | 108  | 3.2  | 1.75  | -1.3 |
| B$_g$ | 109.7 | 2.2  | 1.18  | ---- | ---- | ----  | ---- |
| B$_g$ | 114.4 | 2.0  | 1.03  | 114  | 1.4  | 0.72  | 0.3  |
| B$_g$ | 119.4 | 5.0  | 2.47  |      | ---- | ----  | ---- |
| A$_g$ | 122.7 | 2.5  | 1.20  | 127  | 4.2  | 1.95  | -3.4 |
| A$_g$ | 130.3 | 3.5  | 1.58  | 136  | 2.3  | 1.00  | -4.2 |
| B$_g$ | 155.0 | 5.1  | 1.94  | 144  | 3.3  | 1.35  | 7.6  |
| A$_g$ | 158.7 | 5.3  | 1.97  | 161  | 4.9  | 1.80  | -1.4 |
| B$_g$ | 174.7 | 5.8  | 1.96  | 177  | ---- | ----  | -1.3 |
| B$_g$ | 184.4 | 6.1  | 1.95  | 181  | ---- | ----  | 1.9  |
| B$_g$ | 192.1 | 6.3  | 1.93  | 187  |      | ----  | 2.7  |
| A$_g$ | 192.4 | 4.9  | 1.50  | 196  | 5.0  | 1.51  | -1.8 |
| A$_g$ | 197.4 | 6.7  | 2.00  | 211  |      | ----  | -6.4 |
| B$_g$ | 333.6 | 0.6  | 0.11  | 334  | 0.6  | 0.11  | -0.1 |
| A$_g$ | 344.2 | 1.8  | 0.31  | 342  | 1.7  | 0.29  | 0.6  |
| B$_g$ | 349.0 | 1.3  | 0.22  | 350  | 1.8  | 0.30  | -0.3 |
| A$_g$ | 359.9 | 1.2  | 0.20  | 364  | 1.3  | 0.21  | -1.1 |
| A$_g$ | 367.1 | 2.7  | 0.43  | 367  | ---- | ----  | 0.0  |
| B$_g$ | 389.4 | 2.4  | 0.36  | 376  | 2.6  | 0.41  | 3.6  |
| $A_g$ | 391.8 | 3.7 | 0.56 | 398 | 3.0 | 0.44 | -1.6 |
| $B_g$ | 399.2 | 2.7 | 0.40 | 403 | 3.3 | 0.48 | -0.9 |
| $B_g$ | 423.2 | 1.9 | 0.26 | 424 | 2.0 | 0.28 | -0.2 |
| $A_g$ | 429.1 | 2.3 | 0.32 | 432 | 2.8 | 0.38 | -0.7 |
| $B_g$ | 901.4 | 4.1 | 0.27 | 859 | 5.3 | 0.36 | 4.9 |
| $A_g$ | 904.2 | 4.1 | 0.27 | 868 | 3.9 | 0.27 | 4.2 |
| $A_g$ | 910.2 | 4.4 | 0.29 | 890 | 4.9 | 0.32 | 2.3 |
| $A_g$ | 933.9 | 5.3 | 0.33 | 894 | 4.8 | 0.32 | 4.5 |
| $B_g$ | 938.7 | 4.2 | 0.26 | 918 | 4.4 | 0.28 | 2.2 |
| $A_g$ | 941.1 | 4.6 | 0.29 | 932 | 4.7 | 0.30 | 1.0 |
| $B_g$ | 959.7 | 4.2 | 0.26 | 951 | | | 0.9 |
| $B_g$ | 975.6 | 4.3 | 0.26 | 970 | | | 0.6 |

**Table VI:** Experimental and calculated wavenumbers (ω) determined at 8.9 GPa for Raman modes of scheelite-type $SrCrO_4$ (in $cm^{-1}$) including mode assignment. The pressure coefficients (dω/dP) are also reported (in $cm^{-1}$/GPa) as well as the experimental Grüneisen parameters (γ). The relative difference between measured and calculated frequencies ($R_\omega$) is also given (in %).

| Mode | Theory | | | Experiments | | | $R_\omega$ |
|---|---|---|---|---|---|---|---|
| | ω | dω/dP | γ | ω | dω/dP | γ | |
| $E_g$ | 73.6 | -3.8 | -3.41 | 72 | -2.2 | -1.83 | 2.2 |
| $B_g$ | 124.9 | 0.3 | 0.16 | 127 | 2.9 | 1.37 | -1.6 |
| $E_g$ | 164.8 | 2.7 | 1.08 | 162 | 2.5 | 0.93 | 1.7 |
| $A_g$ | 175.2 | 3.2 | 1.21 | ---- | ---- | ---- | ---- |
| $B_g$ | 219.3 | 5.8 | 1.75 | 221 | 2.9 | 0.79 | -0.8 |
| $E_g$ | 257.5 | 3.7 | 0.95 | 255 | 5.4 | 1.27 | 1.0 |
| $B_g$ | 375.1 | 1.7 | 0.30 | 381 | 2.6 | 0.41 | -1.5 |
| $A_g$ | 375.4 | 2.2 | 0.39 | | | | |
| $B_g$ | 394.4 | 1.2 | 0.20 | 401 | 3.7 | 0.55 | -1.6 |
| $E_g$ | 422.5 | 1.8 | 0.28 | 431 | 3.3 | 0.46 | -2.0 |
| $E_g$ | 932.7 | 4.9 | 0.35 | 888 | 5.5 | 0.37 | 5.0 |
| $A_g$ | 936.9 | 2.9 | 0.20 | 898 | 5.5 | 0.37 | 4.3 |
| $B_g$ | 988.4 | 5.5 | 0.37 | 953 | 5.5 | 0.35 | 3.7 |

**Table VII:** Experimental and calculated wavenumbers (ω) determined at 11.7 GPa for Raman modes of $AgMnO_4$-type $SrCrO_4$ (in cm$^{-1}$) including mode assignment. The pressure coefficients (dω/dP) are also reported (in cm$^{-1}$/GPa) as well as the experimental Grüneisen parameters (γ). The column on the right-hand side shows the relative difference (in %) between calculated and measured wavenumbers.

| Mode | Theory | | | Experiment | | | $R_\omega$ |
|---|---|---|---|---|---|---|---|
| | ω | dω/dP | γ | ω | dω/dP | γ | |
| $B_g$ | 51.6 | -3.2 | -4.28 | 70 | -0.3 | -0.30 | 26.3 |
| $A_g$ | 58.9 | -0.7 | -0.82 | 78 | -0.5 | -0.44 | 24.5 |
| $A_g$ | 97.2 | 2.4 | 1.70 | 82 | 0.6 | 0.50 | -18.5 |
| $B_g$ | 98.8 | 1.4 | 0.98 | 94 | 0.6 | 0.44 | -5.1 |
| $B_g$ | 123.8 | 2.5 | 1.39 | 111 | 2.0 | 1.24 | -11.5 |
| $A_g$ | 124.1 | 2.1 | 1.17 | 116 | 1.3 | 0.77 | -7.0 |
| $A_g$ | 143.5 | 1.2 | 0.58 | 126 | 1.1 | 0.60 | -13.9 |
| $A_g$ | 150.9 | 2.0 | 0.91 | 132 | 1.5 | 0.78 | -14.3 |
| $B_g$ | 163.9 | 2.2 | 0.93 | 141 | 1.5 | 0.73 | -16.2 |
| $A_g$ | 164.5 | 1.9 | 0.80 | 154 | 1.1 | 0.49 | -6.8 |
| $B_g$ | 178.2 | 2.5 | 0.97 | 164 | 2.9 | 1.22 | -8.7 |
| $A_g$ | 188.1 | 2.6 | 0.95 | 175 | 4.1 | 1.62 | -7.5 |
| $B_g$ | 218.6 | 3.7 | 1.17 | 187 | 3.4 | 1.25 | -16.9 |
| $A_g$ | 225.0 | 4.4 | 1.35 | 200 | 3.8 | 1.31 | -12.5 |
| $B_g$ | 233.2 | 4.1 | 1.21 | 226 | 3.4 | 1.04 | -3.2 |
| $B_g$ | 237.6 | 3.6 | 1.05 | 235 | 4.2 | 1.23 | -1.1 |
| $A_g$ | 239.6 | 3.7 | 1.07 | 244 | 4.2 | 1.19 | 1.8 |
| $B_g$ | 269.7 | 4.7 | 1.20 | 262 | 2.2 | 0.58 | -2.9 |
| $A_g$ | 342.5 | 0.8 | 0.16 | ---- | ---- | ---- | ---- |
| $B_g$ | 354.5 | 1.1 | 0.21 | 354 | 1.5 | 0.29 | -0.1 |
| $B_g$ | 372.5 | 1.1 | 0.20 | 371 | 1.0 | 0.19 | -0.4 |
| $A_g$ | 375.6 | 2.0 | 0.37 | ---- | ---- | ---- | ---- |
| $B_g$ | 378.6 | 1.6 | 0.29 | 387 | 1.8 | 0.32 | 2.2 |
| $A_g$ | 390.6 | 2.0 | 0.35 | 395 | 2.8 | 0.49 | 1.1 |
| $B_g$ | 391.0 | 0.8 | 0.14 | 411 | 1.9 | 0.32 | 4.9 |
| $A_g$ | 430.4 | 1.7 | 0.27 | 419 | 1.4 | 0.23 | -2.7 |
| $B_g$ | 439.2 | 2.5 | 0.39 | 434 | 2.2 | 0.35 | -1.2 |
| $A_g$ | 442.7 | 2.7 | 0.42 | 445 | 3.0 | 0.47 | 0.5 |
| ---- | ---- | ---- | ---- | 852 | 2.2 | 0.18 | ---- |
| ---- | ---- | ---- | ---- | 890 | 2.6 | 0.20 | ---- |
| $A_g$ | 926.8 | 2.5 | 0.19 | 904 | 2.9 | 0.22 | -2.5 |
| $B_g$ | 934.3 | 2.3 | 0.17 | 918 | 3.7 | 0.28 | -1.8 |
| $A_g$ | 951.6 | 2.8 | 0.20 | 927 | 3.3 | 0.25 | -2.7 |
| $A_g$ | 964.4 | 3.1 | 0.22 | 933 | 3.7 | 0.27 | -3.4 |
| $B_g$ | 979.8 | 3.0 | 0.21 | 954 | 3.2 | 0.23 | -2.7 |
| $B_g$ | 981.6 | 3.2 | 0.22 | 969 | 2.9 | 0.21 | -1.3 |
| $A_g$ | 997.1 | 3.5 | 0.24 | 996 | 3.3 | 0.23 | -0.1 |
| $B_g$ | 1041.7 | 3.6 | 0.24 | 1002 | 3.6 | 0.25 | -4.0 |

**Figure 1: (color online)** Schematic view of the low-pressure and high-pressure polymorphs of SrCrO$_4$. Sr atoms: green. Cr atoms: blue. Oxygens: red. The coordination polyhedra of Sr and Cr are also shown.

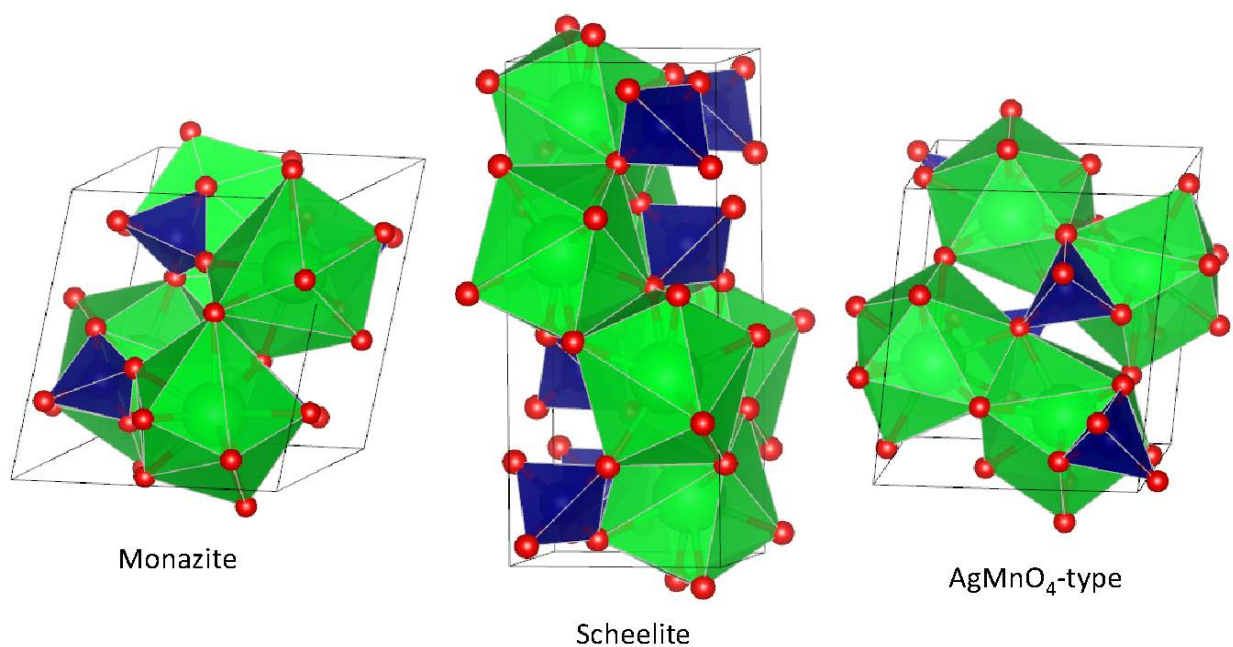

**Figure 2: (color online)** Calculated enthalpy difference versus pressure for the three relevant structures for this study, taking the monazite structure as a reference.

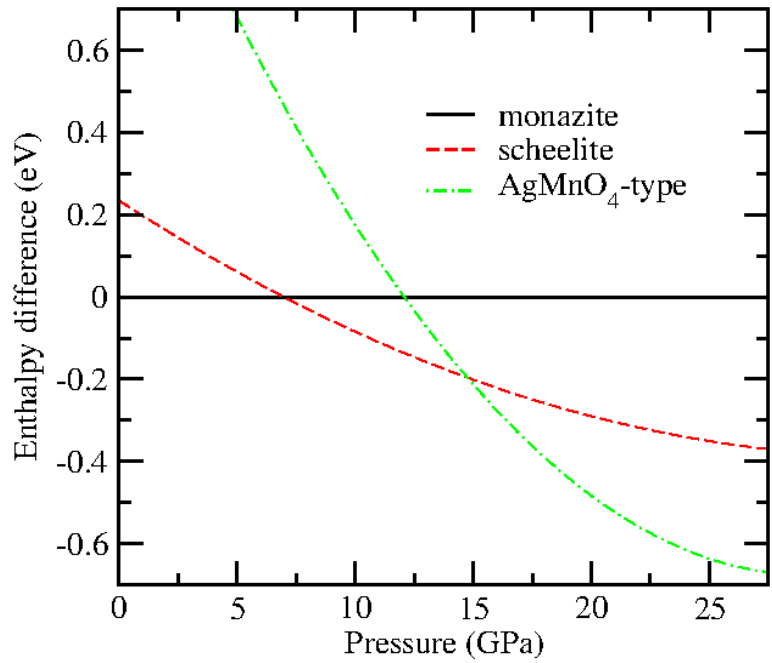

**Figure 3: (color online)** Selection of XRD patterns at different pressures. Dots: experiments (at 20.4 the experiment is shown as a line). Solid lines are the refinements and residuals at all pressure with the exception of 20.4 GPa. Ticks show the position of Bragg reflections. Different colors have been used for the different phases of $SrCrO_4$.

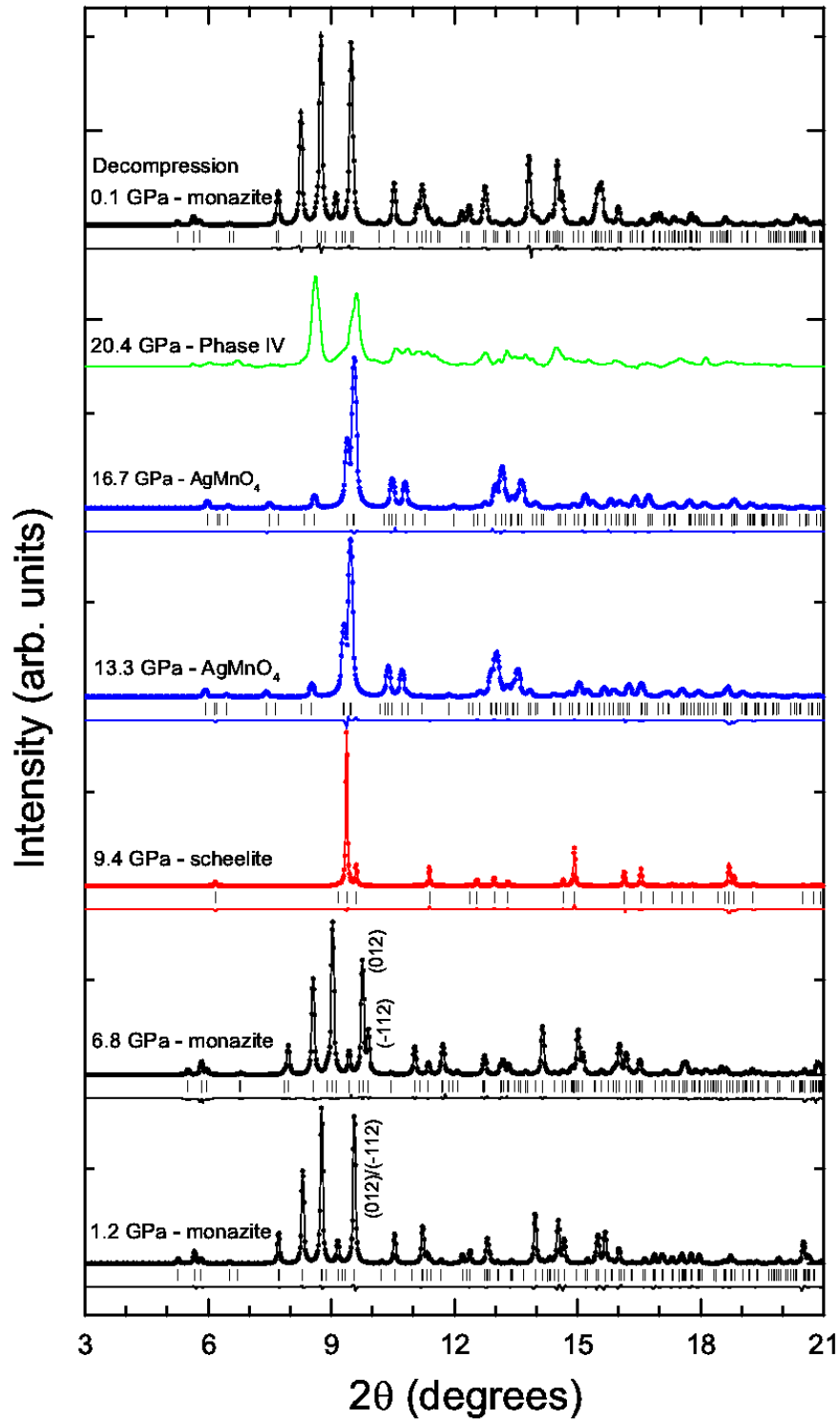

**Figure 4:** Pressure dependence of the unit-cell parameters and volume. Symbols correspond to experiments: Circles: monazite phase. Squares: scheelite phase. Diamonds: AgMnO$_4$ phase. Solid lines show the results of calculations. Error bars are smaller than symbols.

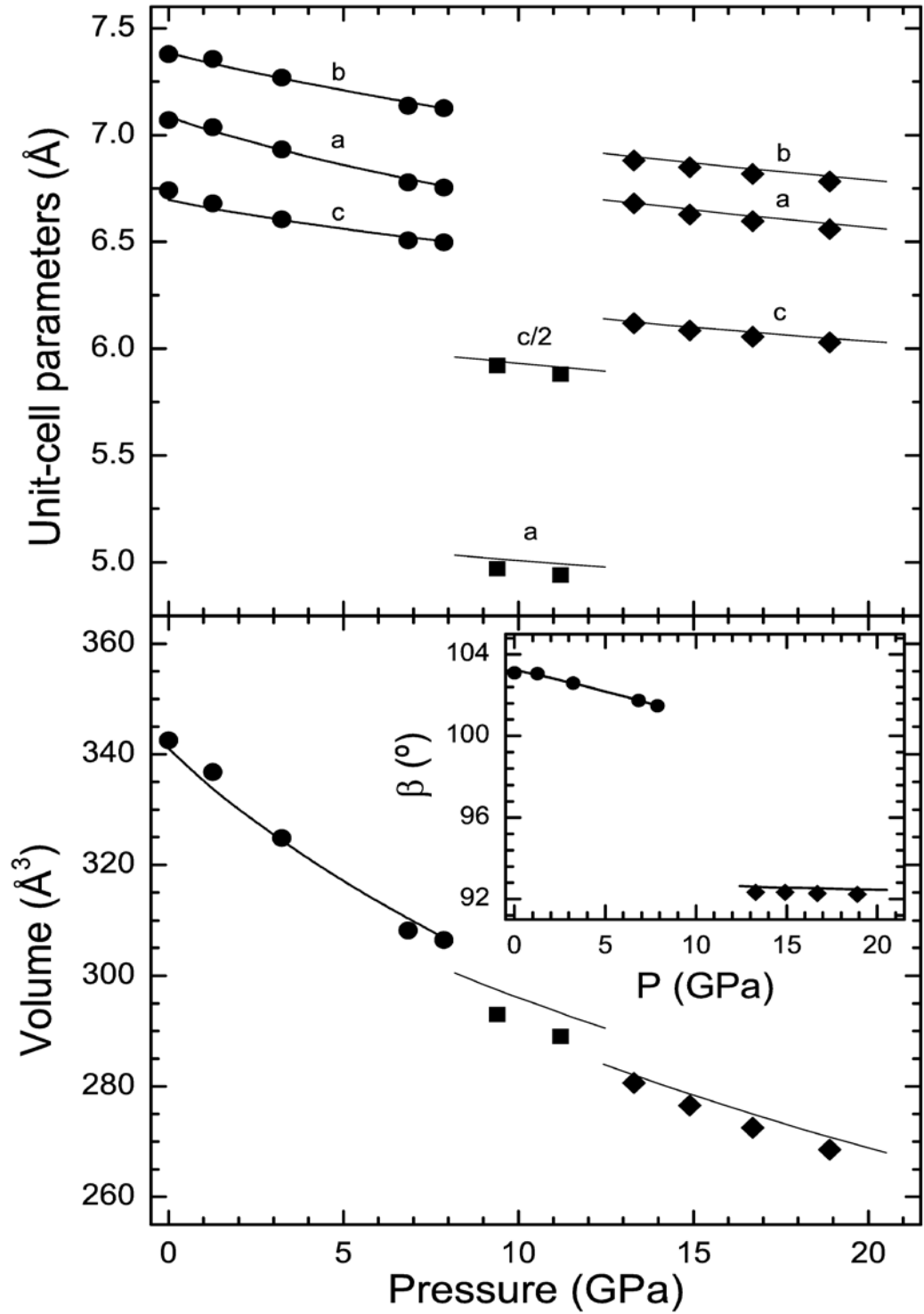

**Figure 5: (color online)** Raman spectra measured at various pressures up to 8.2 GPa using MEW as pressure-transmitting medium. Ticks identify the most intense Raman modes of the monazite-phase.

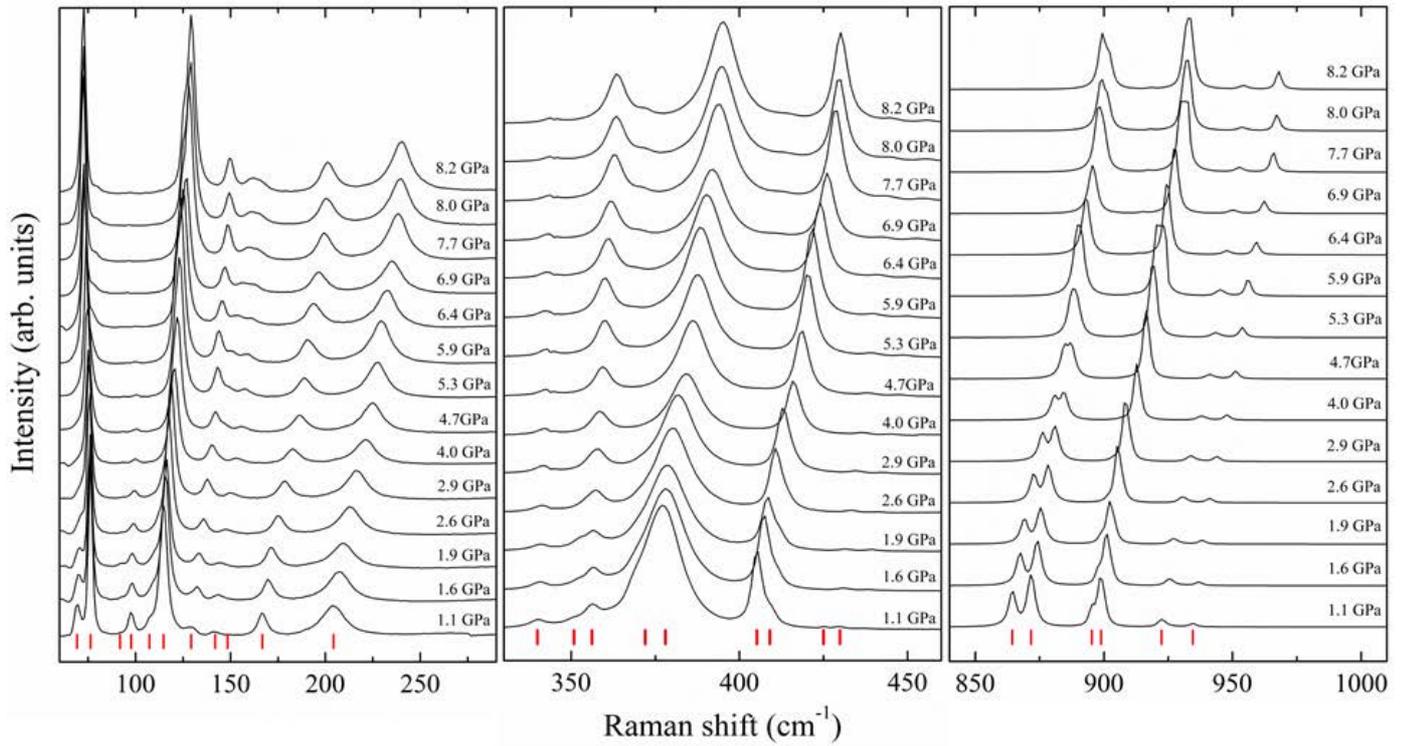

**Figure 7: (color online)** Raman spectra measured at various pressures from 8.9 GPa to 15.7 GPa using MEW as pressure medium. Different phases are identified with different colors. Ticks identify the most intense peaks of the scheelite phase.

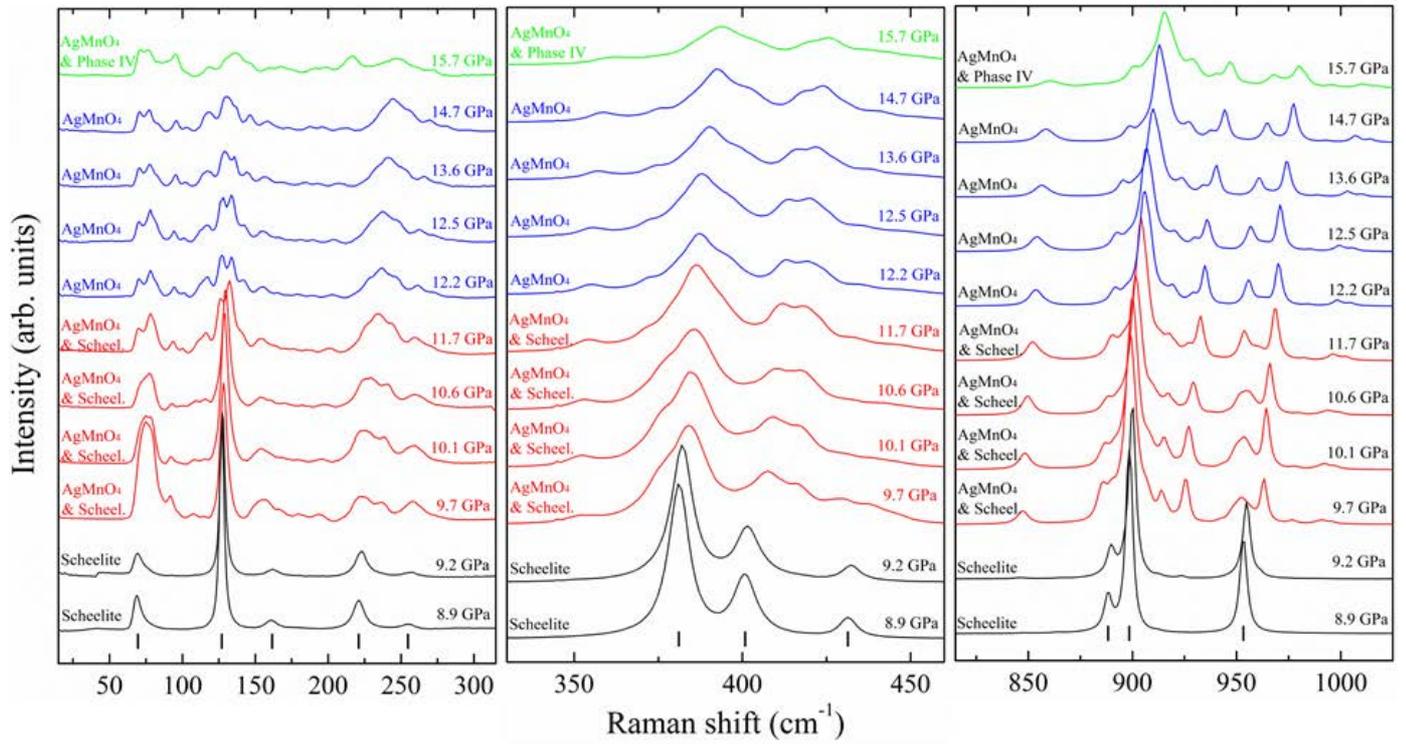

**Figure 7: (color online)** Pressure dependence of the Raman modes of the different phases of SrCrO$_4$. Black squares: monazite phase. Red circles: scheelite phase. Blue diamonds: AgMnO$_4$ phase. Lines correspond to linear fits.

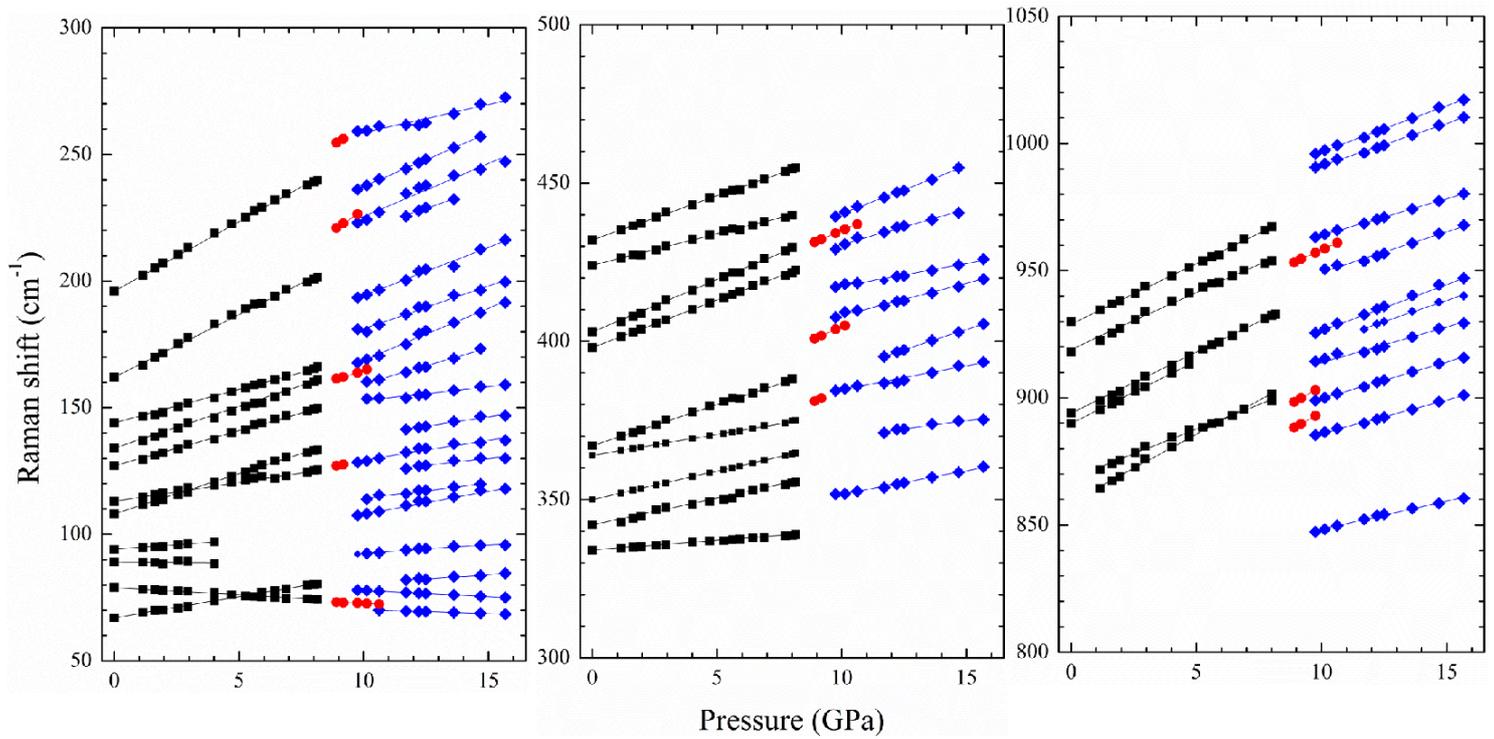

**Figure 8: (color online)** Absorption spectra of SrCrO$_4$ at selected pressures. The arrows point to each band-gap edge.

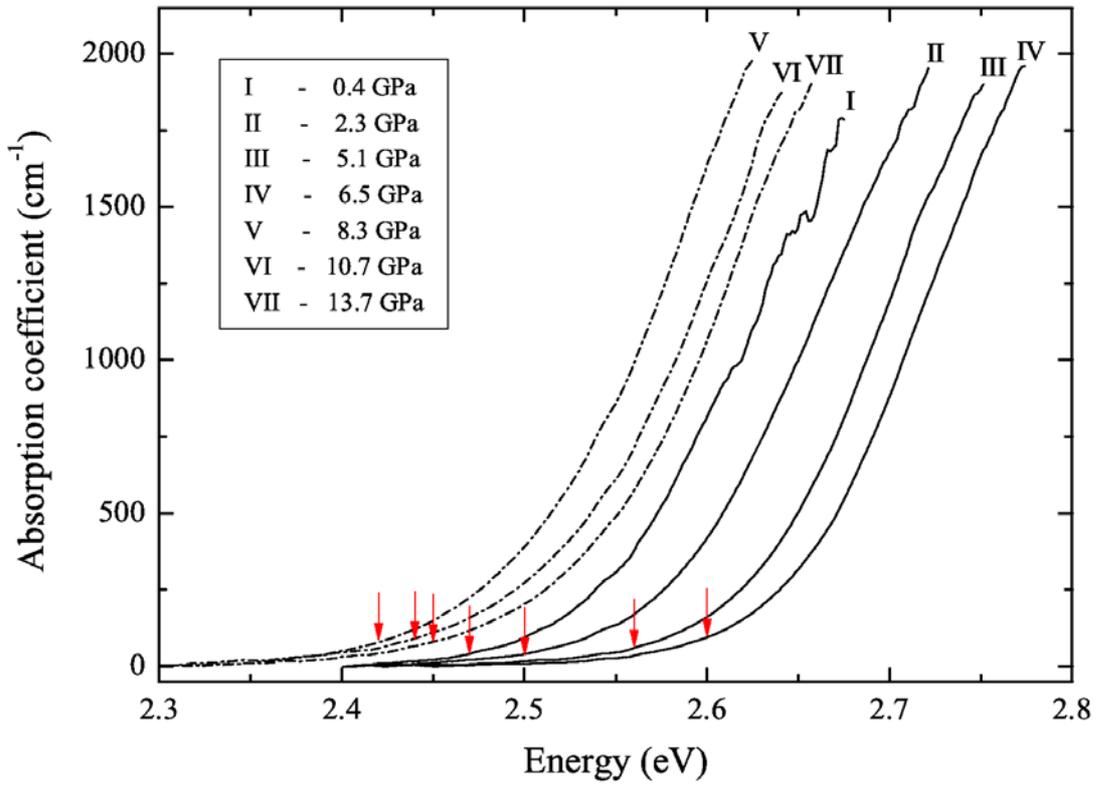

**Figure 9: (color online)** Pressure dependence of the band-gap energy of SrCrO$_4$. Black circles: monazite phase. Red squares: scheelite phase. Blue diamonds: AgMnO$_4$ phase. Solid lines show the results of linear fits.

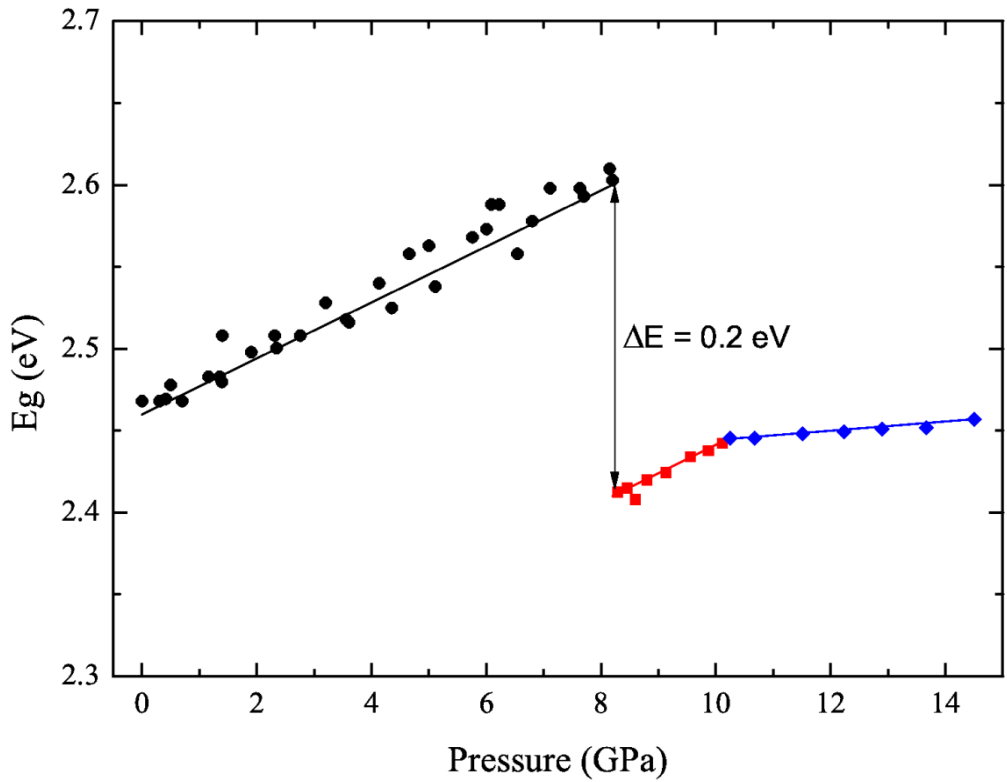